# Investigation of the Yu-Shiba-Rusinov states of a multi-impurity Kondo system

A. Kamlapure[1]*, L. Cornils[1], R. Žitko[2,3], M. Valentyuk[1,4], R. Mozara[1], S. Pradhan[5], J. Fransson[5], A. I. Lichtenstein[1,4], J. Wiebe[1†], and R. Wiesendanger[1]

[1]*Department of Physics, University of Hamburg, Jungiusstrasse 11, D-20355 Hamburg, Germany*
[2]*Jožef Stefan Institute, Jamova 39, SI-1000 Ljubljana, Slovenia*
[3]*Faculty of Mathematics and Physics, University of Ljubljana, Jadranska 19, SI-1000 Ljubljana, Slovenia*
[4]*Department of Theoretical Physics and Applied Mathematics, Ural Federal University, 19 Mira Street, Yekaterinburg, 620002, Russia*
[5]*Department of Physics and Astronomy, Uppsala University, P.O. Box 516, Uppsala SE-751 21, Sweden*

Recent studies of mutually interacting magnetic atoms coupled to a superconductor have gained enormous interest due to the potential realization of topological superconductivity[1–21]. The Kondo exchange coupling $J_K$ of such atoms with the electrons in the superconductor has a pair-breaking effect which produces so-called Yu-Shiba-Rusinov (YSR) states[22–24] within the superconducting energy gap, whose energetic positions are intimately connected with the requirements for topological superconductivity. Here, using the tip of a scanning tunneling microscope, we artificially craft a multi-impurity Kondo system coupled to a superconducting host consisting of an Fe adatom interacting with an assembly of interstitial Fe atoms on an oxygen-reconstructed Ta(100) surface and we experimentally investigate the signatures of Kondo screening and the YSR states. With the help of numerical renormalization group (NRG) calculations, we show that the observed behavior can be qualitatively reproduced by a two-impurity Kondo system whose inter-impurity antiferromagnetic interaction $J$ is adjusted by the number of interstitial Fe atoms in the assembly. When driving the system from the regime of two decoupled Kondo singlets (small $J$) to that of an antiferromagnetic dimer (large $J$), the YSR state shows a characteristic cross-over in its energetic position and particle-hole asymmetry.

* akamlapu@physnet.uni-hamburg.de
† jwiebe@physnet.uni-hamburg.de

## Introduction

Mutually interacting magnetic impurities which are Kondo exchange coupled to the electrons forming the Cooper pairs in a superconducting host constitute building blocks for the realization of a topological superconductor that can potentially host Majorana states on its edges[1–21]. While the Kondo exchange coupling $J_K$ affects the position of the Yu-Shiba-Rusinov (YSR) states inside the gap of the superconductor (2Δ)[22–27], the mutual magnetic interactions $J$ between the impurities dictate how strongly these YSR states split or hybridize into bands[28–34], and govern the collective magnetic state of the magnetic impurity cluster. The investigation of the effects of varying $J_K$ and $J$ on the YSR states in such systems is, thus, of fundamental interest towards the end of realizing topological superconductivity.

If $J_K$ is sufficiently large, the impurity enters the strong-coupling Kondo regime where the impurity spin forms a many-body singlet ground state with the conduction electrons for temperatures below the characteristic Kondo temperature[35], $T_K$. For $k_B T_K \gtrsim \Delta$, the spectroscopic signature of the Kondo screened spin, the so-called Kondo resonance centered at the Fermi energy and having a width of order $k_B T_K$, can coexist with the YSR states[36–38]. For normal metallic substrates, the behavior of two Kondo impurities interacting via the same conduction electron system that produces the Kondo screening, the so-called two-impurity Kondo problem[39,40], has been studied in great detail[41–45]. For antiferromagnetic (AFM) inter-impurity interaction $J$ of increasing strength, the system passes over from two decoupled Kondo singlets (small $J$) to an AFM dimer (large $J$), which causes the Kondo resonance to split. However, the analog of this two-impurity Kondo problem for superconducting host systems[31,46–48] has not been studied experimentally to a large extent so far[49,50], and it is an interesting question, how the YSR states behave as a function of $J$.

In this work, we experimentally study artificial Fe clusters consisting of a single Fe adatom magnetically interacting with an assembly of interstitial Fe atoms (IFA) of different shapes and sizes on an oxygen reconstructed superconducting Ta(100) surface. Through comparison with NRG calculations, our spectroscopic investigation shows that the system qualitatively behaves as a two-impurity spin-1/2 Kondo system coupled to a superconductor with $k_B T_K \sim J \gg \Delta$, where the first Kondo impurity is associated to the Fe adatom and the second to the IFA assembly. As a function of the size of the IFA assembly, the system shows a smooth transition between the weak and strong exchange coupling regimes (i.e., independent Kondo singlets vs. magnetically aligned spins) observed through the formation of a spectral depletion in the Kondo peak, i.e., an exchange-splitting gap. Such simple behavior is surprising, since the impurities have multiple orbitals, potentially producing high-spin states. Furthermore, it is not *a-priori* expected that multiple IFAs can be conflated into a single effective spin degree of freedom. Interestingly, simultaneously with the transition observed in the shape of the Kondo peak on the large energy scale of $k_B T_K \sim J$, the YSR state energy whose scale is small compared to $\Delta$, as well as the particle-hole asymmetry of this state, reveal a characteristic crossover that can likewise be assigned to the variation of $J$ with the help of the NRG calculations.

## Results

### Construction of the Fe clusters

The details of the preparation of a (3 x 3) oxygen-reconstructed Ta(100) surface, in the following called TaO, and of the cold-deposition of Fe can be found in Refs. [4,51,52]. We use vertical atom manipulation to build various clusters consisting of a single Fe adatom and an assembly with different numbers of IFAs in close proximity (see Methods). Figure 1a shows a representative surface of TaO where an Fe adatom was placed at the center of the (3 x 3) plaquette and a single interstitial Fe atom was positioned between two (3 x 3) plaquettes[4]. Figure 1c shows the result of a geometry optimization of the surface through DFT calculations[52] revealing the stable positions of the Fe adatom and the interstitial Fe atom. Within the DFT calculations, we find a finite magnetic moment for both the Fe adatom as well as the IFA (see methods). The schematic top views and STM images of the other investigated clusters with 0 – 4 IFAs are shown in Figs. 1d,e (see Supplementary Fig. la for other examples of similar clusters). To build such clusters, initially, we moved Fe atoms to the interstitial locations to form the assembly, followed by placing the Fe adatom in close proximity (case 1 – 4). We also studied the single Fe adatom without an assembly of IFAs for comparison (case 0).

**Spectroscopy of the Fe clusters**

Spectra measured on these clusters with a superconducting tip in a wide bias range (Fig. 2a) show a resonance around the Fermi energy. Here we have used a large amplitude of the modulation voltage such that the superconducting gap and features within it are washed out. For the bare adatom, the resonance is very broad (half-width at half maximum (HWHM) $\Gamma\sim$ 20-25 mV). The adatom in proximity to one interstitial Fe atom shows instead a rather sharp resonance at the Fermi level [$\Gamma\sim 12\ (\pm 1)$ mV]. We therefore assume that the resonance is of many-body origin (i.e., due to the Kondo effect), and interpret the differences of the spectral width of the resonance as an effect of the neighboring atoms on the Fe adatom's Kondo coupling $J_K$ to the substrate electrons. Moreover, it is remarkable to see that, when the adatom is coupled to an assembly with two or more IFAs, the resonance peak appears to split, resulting in a zero bias anomaly (ZBA). A careful look at the spectra for adatoms showing ZBA reveals that, during the formation of the gap-like structure at the Fermi level, the overall width of the Kondo resonance remains unchanged. More precisely, the asymptotic tails of the peaks in the sequence of 1,2,3,4 IFAs overlap to a good approximation, indicating that $J_K$ remains constant, while an increasingly wide gap is forming at the peak center. In order to quantify the width of the zero bias resonance $\Gamma$ and the gap width $E_s$, we fit each dI/dV spectrum in Fig. 2a to a Fano curve[53] with an additional multiplicative factor featuring a Lorentzian dip for the spectra with a ZBA (see Methods and dashed lines in Fig. 2a). It is evident from the fits that the details of the spectral shape are excellently captured using this functional form. Similar fits performed for a number of other clusters are shown in Supplementary Fig. 1b. The Kondo temperatures $T_K$ extracted from the fitted values of $\Gamma$ (see Methods) are on the order of 25 K - 40 K. It is interesting to note that the Kondo energy scale in our system is very large compared to the superconducting gap[4,51] ($k_B T_K \gg \Delta$), and that the width of the ZBA, $E_S$, is of similar size as $k_B T_K$. We will show below that the ZBA can be interpreted as an exchange-splitting gap due to a strong interaction between the Fe adatom and the assembly of IFAs.

Next, dI/dV spectra taken on the same clusters in the low-bias range with small ac modulation will be discussed (Fig. 2b). They reveal sharp peaks inside the energy gap of the superconducting substrate due to YSR states[4,51]. Since the measurement is performed using a superconducting tip, we extract the corresponding local electron density of states (LDOS) using a standard deconvolution method[4,36]. The corresponding curves (Fig. 2c) show two sharp peaks, one with larger intensity located at $E_{\mathrm{YSR}}$ and one with smaller intensity located at -$E_{\mathrm{YSR}}$. We observe a large variation of $E_{\mathrm{YSR}}$ as the Fe cluster size and its configuration changes. In particular, $E_{\mathrm{YSR}}$ tends to decrease and reverse its sign when the number of Fe atoms in the assembly is increased (see, in particular, configurations 1, 2, and 3). This is also apparent from the change in the sign of the asymmetry in the intensities of the two peaks (see Supplementary Fig. 2 for the definition of the asymmetry and data of all clusters). In order to establish a possible correlation between the YSR states and the shape of the Kondo resonance, we plot $E_{\mathrm{YSR}}$ normalized to $\Delta$ as a function of the reduced Kondo temperature, $k_{\mathrm{B}} T_{\mathrm{K}}/\Delta$ (Fig. 3a) and, as a function of $E_{\mathrm{S}}$ (Fig. 3b) extracted from the fitting.

**Correlation between Kondo resonance width and YSR energy**

From Fig. 3a, it is evident that the Kondo temperature mostly changes by a factor of 1/2 (from $(k_B T_K/\Delta) \sim 10$ to 5) when a single IFA is added to the Fe adatom, but does not further reduce with the number or configuration of IFAs in the assembly. At the same time, $E_{\mathrm{YSR}}$ strongly changes from $\sim\Delta$ to $-\Delta$ when the number of IFAs in the assembly is increased. Consequently, there is almost no correlation between $E_{YSR}$ and $T_K$. At first glance, this result seems surprising, as similar trends of experimentally observed YSR energies were usually assigned to the quantum phase transition of the system induced by an increasing Kondo coupling[25–27,36–38]. In order to exclude this effect as the major reason for the observed changes to $E_{YSR}$, we simulated spectra within Slave-boson mean-field theory (SBMFT), which is ideally suited for the strong-coupling Kondo regime $[(k_B T_K/\Delta) \gg 1]$ and thus valid for the experimentally investigated system[54–56]. We consider a simple model consisting of a single spin-1/2 Anderson impurity coupled to a superconducting substrate (see Methods). The SBMFT simulation of $E_{YSR}$ as a function of the Kondo temperature is plotted as a violet line in Fig. 3a. The deviation of the experimental data points from the simulation infers that the trend in $E_{YSR}$ in our case is not governed by changes to the Kondo coupling alone. This is also evident from the comparison of our data to experimental data and NRG calculations for another system[37] (Supplementary Fig. 3), where we see that the system of Fe clusters on TaO investigated here always stays on one side of the Kondo-coupling induced quantum phase transition. Only the initial decrease of $T_K$ by a factor of 2 when a single IFA is added to the Fe adatom is the result of a change of its Kondo coupling, presumably through a small structural relaxation.

On the other hand, from Fig. 3b, we observe that $E_{YSR}$ is strongly correlated with the ZBA in the large-bias spectra. In particular, while the YSR state energy is positive for most of the Fe clusters with a small number of IFAs in the assembly which do not show a ZBA in the Kondo resonance ($E_S = 0$), it is negative for those with a large number of IFAs which have a non-zero $E_S$. The ZBA of the Kondo resonance can be naturally interpreted as an exchange splitting induced by the competition of Kondo screening ($T_K$) and exchange coupling of the Fe adatom with the assembly of IFAs ($J$) which probably changes by the number of IFAs in the assembly[41,42,45]. In order to qualitatively test this hypothesis, we consider the following minimal model. We regard our system as a realization of the two-impurity Kondo problem where the assembly of the IFAs is represented by a single effective spin-1/2 impurity interacting via exchange interaction $J$ with the adatom as a second spin-1/2 impurity. Within this model, for small AFM interaction, the two spins are independently screened by the conduction electrons resulting in a single peak in the spectral function. For very strong interaction ($J \geq J_c$), the two spins undergo a crossover to a local AFM singlet state and are no longer screened independently [41,42,45,57–59]. The spectral manifestation of this phenomenon is the observed splitting of the Kondo peak which for small and moderately large values of $J$ takes the appearance of a "spin-gap" of width $E_S$ inside the resonance. The reason behind this particular line-shape is that the Kondo screening proceeds unperturbed at high energy scales, until it reaches the scale of $J$ where the residual magnetic moments rapidly bind antiferromagnetically. This matches the evolution of the Kondo peak shape observed in the experiment. The width of the gap $E_S$ seen in the spectra with ZBA is, therefore, a direct measure of the effective strength $J$ of the exchange interaction between the adatom and the assembly of the IFAs taken as a whole. Moreover, the order of magnitude of $E_S$ of 2 meV - 8 meV, implies a large AFM interaction between the adatom and the assembly of IFAs. The hierarchy of parameters is thus $J \sim T_K \gg \Delta$ [41,45,57].

**NRG calculations**

In order to check whether the competition of Kondo screening ($T_K$) and exchange coupling $J$ within this minimal model can also explain the experimentally observed correlation between the width of the exchange-splitting gap $E_S$ and the YSR state binding energy $E_{YSR}$ (Fig. 3b), we have performed NRG calculations that produce reliable spectral functions on all energy scales, including the region inside the superconducting gap[60,61]. We have used a two-impurity Anderson model to describe the system (see Methods section), with one impurity orbital representing the IFA assembly, and one representing the Fe adatom chosen to be particle-hole asymmetric. This model can also account for the effects of potential scattering on the adatom. The results show that while the ground state of the system is a spin singlet for all values of exchange coupling ($J$), its nature smoothly changes from a Kondo singlet to an AFM singlet as indicated e.g. by the increasing magnitude of the spin-spin correlation function. This is reflected by the emergence of the Kondo peak splitting for $J \sim J_c = 5\Delta$ ($J_c$ being the critical exchange interaction) (Fig. 4a) and the characteristic evolution of the YSR peaks (Fig. 4b). To understand the latter, we remind[46] that for two decoupled impurities, the sub-gap spectrum consists of a singlet (Kondo) ground state $S_1$, two doublet excitations $D_{1,2}$ (unscreened impurity states of either impurity), as well as a degenerate singlet $S_2$ and triplet $T$, which correspond to the excited states where both impurities are unscreened. With increasing $J$, the singlet $S_2$ decreases in energy, until it mixes with the Kondo state $S_1$ leading to an anti-crossing shape in the energy diagram. For large $J$, the Kondo state $S_1$ is pushed to higher energies together with the two doublet excitations $D_{1,2}$, while $S_2$ is the new ground state. This leads to the observed evolution of YSR peaks in Fig. 4b, which for small $J$ slightly decrease in energy, then for large $J$ rapidly increase in energy and eventually merge with the continuum (Fig. 4c). The change of direction of YSR shifting occurs at $J \sim J_c$ which is very close to $J$ where the exchange splitting becomes observable in the Kondo peak (compare Fig. 4c with 4d). This behavior is in qualitative agreement with the trend observed in the experiment, as shown by the comparison of the plot of $E_{YSR}$ versus $E_S$ extracted from the NRG-calculated spectra with the same parameters extracted from the experimental data (see Fig. 3b). We also observe that the sign change in the asymmetry of the YSR peak intensities occurs across $J \sim J_c$, in line with the experiment (see Supplementary Fig. 2).

## Conclusions

In conclusion, the experimentally observed correlation between the small energy scale position of the YSR state and the large energy scale exchange-splitting gap of the Fe adatom coupled to the assembly of IFAs can be assigned to changes in the effective exchange coupling $J$ between the adatom and the assembly in the regime where $J \lesseqgtr T_K \gg \Delta$. We have shown that the Fe adatom coupled to a single IFA has a reduced Kondo coupling with respect to the isolated Fe adatom, while further addition of IFAs to the assembly only increases $J$. We compared our experimental results to NRG calculations employing a minimal model that shows how with increasing exchange interaction the system of Fe clusters undergoes a crossover from the regime of an independent Kondo cloud to a local magnetically aligned state. This crossover results in the opening of a gap in the Kondo peak and in the anti-crossing of the sub-gap YSR states. Our results also indicate that introducing the potential scattering results in the change of the particle-hole asymmetry of the YSR peaks across this crossover, thereby facilitating its observation.

Our work establishes the important role of exchange interaction on the experimentally observed shifts in YSR peaks, which were previously supposed to be prevalently determined by the Kondo coupling of the impurity with the substrate. Our results, therefore, challenge the current understanding of the intimate correlation between Kondo effect and YSR state and motivate further studies. In particular, it is becoming clear that while a phenomenological description in terms of hybridizing classical YSR states is possible and valid, the calculation of its parameters should be based on a microscopic quantum model that needs to properly incorporate exchange coupling (including its transverse parts that correspond to spin-flip terms). This calls for further development of efficient theoretical tools for solving sizable assemblies of quantum spins in interaction with a superconducting environment.

## Methods

### Experimental methods:

All the experiments were carried out in a custom-built commercial cryostat (SPECS) at a temperature of $T$ = 1.2 K in ultra-high vacuum. Details of the sample cleaning and cold Fe deposition can be found in Refs. [4,51]. To make different clusters of Fe atoms, we use vertical atom manipulation where Fe atoms were first transferred from the sample surface to the tip and then dropped back to the desired location. Details of the vertical atom manipulation and also the interstitial atom manipulation are described in Ref. [4]. For spectroscopy, we used a superconducting tip which was prepared by coating a Cr tip with Ta via a controlled dipping of the tip into the Ta substrate. The d$I$/d$V$ spectra were measured with the standard lock-in technique with a modulation frequency of $f$ = 827 Hz and a typical modulation voltage of $V_{\mathrm{mod}}$ = 20 μV (2 - 3 mV) for low (high) bias range spectra. To extract the local density of states from the measured spectra in the low-bias range, we use a standard deconvolution technique [4,36].

### Fitting of the Kondo resonance

In order to determine the Kondo temperature, we fit d$I$/d$V$ spectra taken on the adatom to a Fano function given by[53]:

$$\sigma(V) \propto \sigma_0 + \sigma_1 \frac{\left(q + \frac{V - V_0}{\Gamma}\right)^2}{1 + \left(\frac{V - V_0}{\Gamma}\right)^2}$$

where $q$ is the form factor which determines the line shape of the spectra, $\Gamma$ is the half width at half maximum (HWHM), and $V_0$ is the bias value at the maximum of the spectra. $\sigma_0$ and $\sigma_1$ are the offset and the amplitude of the spectra, respectively.

For those spectra showing the resonance and a ZBA, we fit the dI/dV curves to the product of a Fano function and an inverted Lorentzian, given by:

$$\sigma(V) \propto \sigma_0 + \sigma_1 \left[ \frac{\left(q + \frac{V - V_0}{\Gamma}\right)^2}{1 + \left(\frac{V - V_0}{\Gamma}\right)^2} \right] \times \left[ 1 - \frac{\sigma_2}{1 + \left(\frac{V - V_s}{\Gamma_s}\right)^2} \right]$$

Here, $E_s = e\Gamma_s$ is the half width in energy of the ZBA gap, $V_s$ is the position of the gap, and $\sigma_2$ is a constant.

Finally, we use the HWHM, $\Gamma$, to extract the Kondo temperature of each assembly and use Wilson's definition[62] to define the Kondo temperature as $T_K = 0.27e\Gamma/k_B$, where $k_B$ is the Boltzmann constant.

## Theoretical methods:

## DFT method:

To determine the low-temperature composition of the Fe cluster with IFA, we performed a set of DFT simulations using the VASP package[63,64]. A minimal supercell of the bcc(001) crystal was built from five layers of Ta atoms with a lattice constant of 3.30 Å. The supercell of the surface was chosen to reproduce the area of one (3 x 3) plaquette, as seen in STM images. All simulations were done using spin-polarized functionals using the generalized gradient approximation (GGA+U) for the exchange-correlation part[65,66]. The standard calculation was performed with $U$ = 4.3 eV, $J$ = 0.9 eV for the Fe $d$-states[67] and $U$ = 6 eV, $J$ = 0.8 eV for the O $p$-states. An enlarged cut-off energy was set up to 500 eV and the number of bands was increased by 50%. The structural optimization was performed until forces were less than 0.01 eV/Å on the 6 × 6 × 1 $k$-points grid. We searched for the possible positions of IFAs in the first interspace of the TaO substrate, i.e. between the first and the second layer of the bcc(001)-arranged Ta atoms. These configurations of the surface with a single IFA were then compared by total energies and the induced type of reconstructions (see Supplementary Fig. 4).

The adsorption of the Fe adatom at the center of the (3 x 3) plaquette was studied separately to account for van-der-Waals (vdW) contributions at the surface. The slab was transformed to the symmetric setup of five layers with two identical surfaces to avoid an interaction with charge images. The midpoint of the (3 x 3) plaquette coincides with the central Ta atom, which is marked as a particularly polarized atom in the surface electronic structure[52]. To account for the polarizability of the local structures on the surface, we performed a set of structural optimizations including dispersion correction energies, with an accuracy of atomic forces of less than 0.001 eV/Å. First, the slab with an adatom in the central position was relaxed using the Tkatchenko-Scheffler (TS) method[68], where all surface atoms, including the adatom, were free to relax. To account for possible corrections from the ionic states of the surface, we also included the Hirschfield iterative (HI)[69,70] partitioning scheme (TS/HI) with a self-consistent cycle[71,72]. After relaxation, the Fe adatom has a tiny in-plane offset from the symmetric central position, with a magnitude of the vector of only 0.001 Å. The extracted position of the adatom was then used to compare between different methods in simulations with both the Fe adatom and the IFA.

## Numerical renormalization group method:

We have used the "NRG Ljubljana" implementation of the numerical renormalization group. The calculations have been performed for a two-impurity Anderson model given by:

$$H_\alpha = \sum_{k\sigma} \epsilon_k c^\dagger_{k\sigma\alpha} c_{k\sigma\alpha} - \Delta \sum_k \left( c^\dagger_{k\uparrow\alpha} c^\dagger_{-k\downarrow\alpha} + H.c. \right) + \sum_\sigma \epsilon_\alpha d^\dagger_{\sigma\alpha} d_{\sigma\alpha} + U n_{\uparrow\alpha} n_{\downarrow\alpha} + \sum_{k\sigma} V_{k\alpha} \left( d^\dagger_\alpha c_{k\alpha} + H.c. \right)$$

$$H_{12} = J s_1 \cdot s_2 - t \sum_\sigma \left( d^\dagger_{1\sigma} d_{2\sigma} + H.c. \right)$$

$$H = H_1 + H_2 + H_{12}.$$

Here $\alpha$ indexes the two impurities ($\alpha$ = 1, 2), $\sigma$ is the spin (↑, ↓), $k$ is the electron momentum, $\epsilon_k$ the dispersion of band electrons, and $V_{k\alpha}$ is the hybridization between the impurity $\alpha$ and the superconducting band ($\Gamma = \pi\rho|V|^2$, where $\rho$ is the density of states in the band in the absence of superconductivity). Finally, $n_{\sigma\alpha} = d_{\sigma\alpha}^{\dagger} d_{\sigma\alpha}$ is the occupancy operator for impurity $\alpha$, while $s_\alpha$ is the impurity spin operator defined as $s_\alpha = (1/2) d_{i\alpha}^{\dagger} \boldsymbol{\sigma}_{ij} d_{j\alpha}$, where $\boldsymbol{\sigma}$ is the vector of Pauli matrices, while *i,j* are the internal spin indexes.

We use the parameters $U_1 = U_2 = 10$, $\Gamma_1 = \Gamma_2 = 0.7$, $\epsilon_1 = -4$, $\epsilon_1 = -5$, with the gap fixed at $\Delta = 0.001$ (all parameters are expressed in units of the bandwidth). We have included a small constant hybridization between the impurities ($t = 0.02$) to suppress the sharp quantum phase transition; this is the main motivation for modelling in terms of the Anderson model instead of the pure-spin Kondo model. The overall strength of the exchange coupling is controlled by explicitly including a variable exchange coupling *J*. The calculations were performed for the discretization parameter $\Lambda = 2$ keeping up to 2000 state multiplets and averaging over $N_z = 4$ interleaved discretization meshes. For studying the shape of the Kondo resonance, we have broadened the results in order to smooth out the superconducting gap. This is achieved by using the kernel that corresponds to the sine modulation used in the lock-in technique.

## Slave-boson mean-field theory calculations:

The total Hamiltonian for an Anderson impurity coupled to a superconducting substrate is $H = H_S + H_D + H_T$. Here the substrate is described by the Hamiltonian $H_S = \sum_{k\sigma} \epsilon_{k\sigma} c_{k\sigma}^{\dagger} c_{k\sigma} + \triangle \sum_{k\sigma} c_{k\uparrow}^{\dagger} c_{-k\downarrow}^{\dagger}$, where we assume that the substrate density of states in the normal state is constant within a bandwidth of 2D and $\triangle$ is the energy gap of the superconductor. The tunneling Hamiltonian is given by $H_T = \sum_{k\sigma} v_k (c_{k\sigma}^{\dagger} d_{i\sigma} + h.c.)$ and the tunneling amplitude is given by $\Gamma_i(\omega) = \pi \sum_k |v_{ik}|^2 \delta(\omega - \epsilon_k)$. Here we drop the energy dependency of the tunneling amplitude and assume that it is constant over the energy window $\Gamma_i(\omega) = \Gamma_i = 0.016D$. The Hamiltonian for the impurity is $H_D = \sum_\sigma \epsilon_d d_\sigma^\dagger d_\sigma$, where $\epsilon_d$ is the energy level of the individual impurity. We solve this model using Slave-boson mean-field theory (SBMFT)[35,54,55,73–75]. We restrict the Hilbert space such that there is no double occupancy in the impurity due to a large Coulomb repulsion. This is done by introducing a bosonic and fermionic degree of freedom for each electronic operator, namely $b_{i\sigma}^\dagger = f_{i\sigma}^\dagger b_i$, where $f_{i\sigma}$ is fermion and $b_i$ is boson creation operator. No double occupancy is achieved by the constraint $b_i^\dagger b_i + \sum_{i\sigma} f_{i\sigma}^\dagger f_{i\sigma} = 1$. Within the mean-field approximation, the bosonic operator is replaced by a number and the constraint is fulfilled on an average by introducing a chemical potential.

In the large $T_K$ limit it is favorable to break the Cooper pair such that the impurity spin can be screened. SBMFT is supposed to be valid in this regime[55,56], and thus, in particular, for our experimental setup $k_B T_K/\Delta >> 1$. The spectrum of a single magnetic impurity in a superconductor consists of a continuum at higher energy and a bound state within the gap. This bound state appears close to the Fermi energy for small $T_K$ and shifts towards higher energy for larger $T_K$. Figure 3a shows the evolution of bound states for different values of $T_K$.

## Acknowledgements

The authors thank Markus Ternes for fruitful discussions. This work was supported by the European Research Council Advanced Grant ADMIRE (project no. 786020). R.W. and J.W. acknowledge funding by the Cluster of Excellence 'Advanced Imaging of Matter' (EXC 2056 - project ID 390715994) of the Deutsche Forschungsgemeinschaft (DFG). R.Ž. acknowledges support by the Slovenian Research Agency (ARRS) under Program No. P1-0044. M.V. and R.M. acknowledge financial support from the Deutsche Forschungsgemeinschaft (DFG) through Project No. SFB668-A3 and A1, and from the European Union's Horizon2020 research and innovation program under grant agreement No. 696656 - GrapheneCore1. The work of M.V. was additionally funded through 2019 Equal Opportunity Fund of University of Hamburg. The DFT computations were performed with resources provided by the North-German Supercomputing Alliance (HLRN). S.P. and J.F thank Stiftelsen Olle Engqvist Byggmästare and Vetenskapsrådet for financial support.

## Author contributions

A.K., L.C. and J.W. conceived and designed the experiments. A.K. and L.C. performed the experiments. A.K., J.W. and R.Ž. analyzed the experimental data. NRG calculations were performed by R.Ž.. M.V. and R.M. carried out the DFT calculations. S.P. and J.F. performed SBMFT calculations. A.K. wrote the manuscript. All authors discussed the results and commented on the manuscript.

## Figures

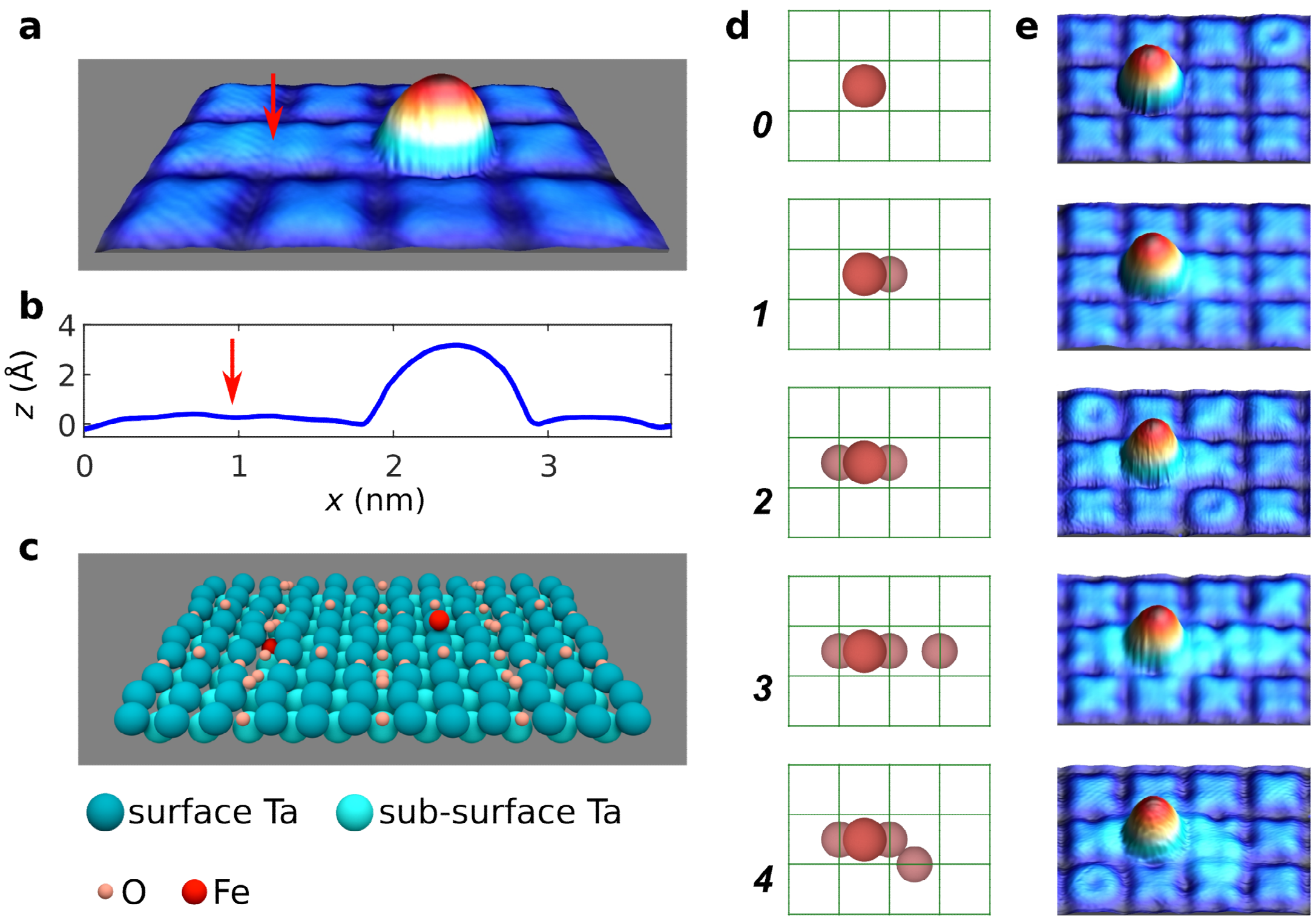

**Figure 1 | Fe adatom and interstitial Fe assemblies at the TaO surface. a**, Constant-current STM image of the TaO surface with an Fe adatom and an interstitial Fe atom (IFA) indicated by the red arrow. **b**, Height profile across the IFA and the Fe adatom in **a**. **c,** Perspective view of the DFT-optimized fully relaxed positions of an Fe adatom, an IFA, and the atoms in the TaO surface. **d**,**e**, Schematic top-view (**d**) and the corresponding STM images (**e**) of an Fe adatom with assemblies of the given different numbers of IFAs ($V$ = 50 mV, $I$ = 100 pA). In **d**, the green lines represent the rows of oxygen atoms, and the larger (smaller) spheres represent the adatom (the IFAs).

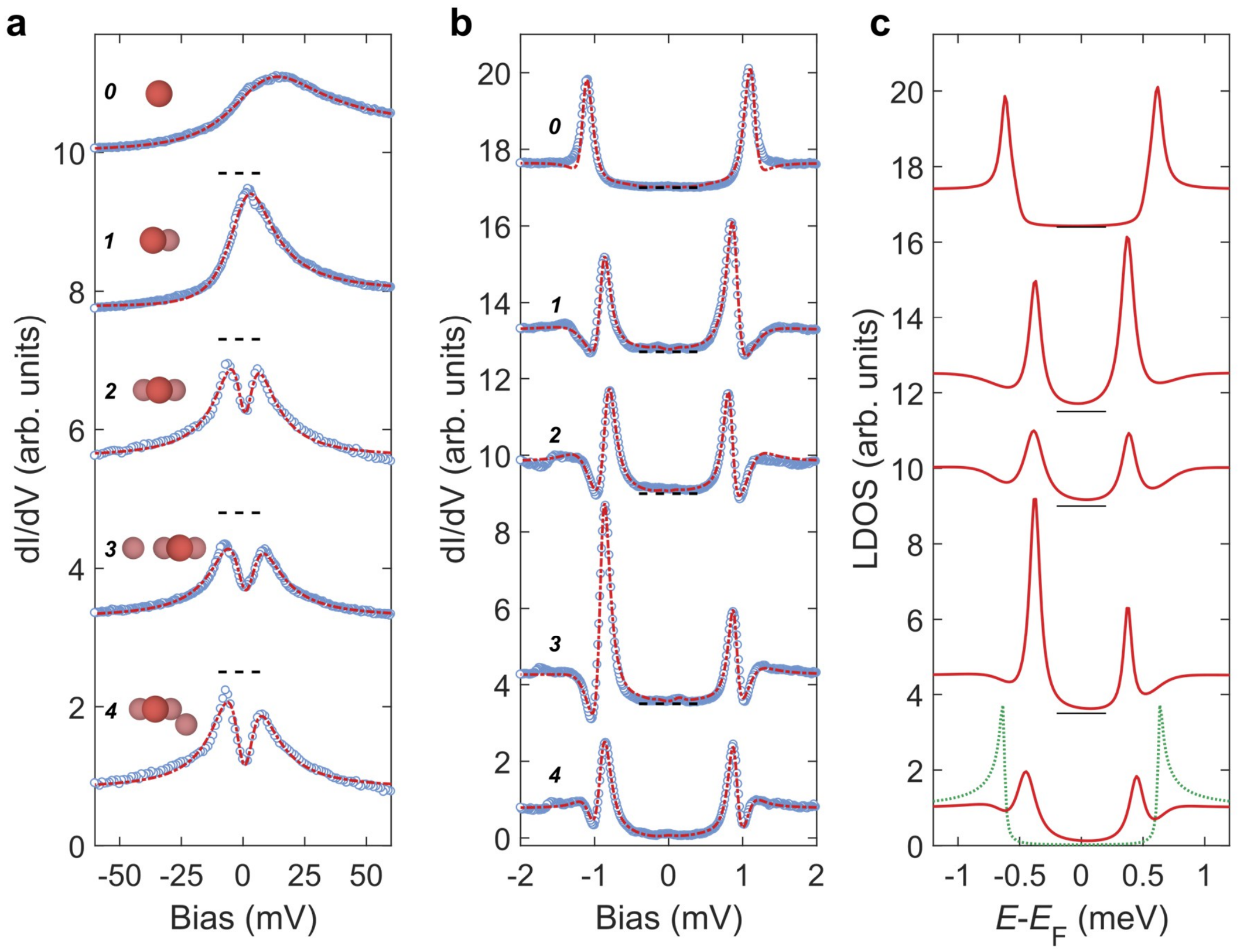

**Figure 2 | dI/dV spectroscopy on Fe adatoms for various interstitial Fe assemblies. a**, Large-bias dI/dV spectra (open dots) showing the Kondo resonance. The dashed lines are fits to a Fano function, which is multiplied by a Lorentzian dip for the spectra exhibiting the zero-bias anomaly (ZBA), i.e. the exchange splitting of the Kondo resonance (see Methods). The pictograms and numbers on the left side of the spectra show the corresponding structure and numbers of the IFA assemblies. Spectra are shifted vertically for clarity. Stabilization parameters for STS: $V_{stab}$ = 100 mV, $I_{stab}$ = 100 pA, $V_{mod}$ = 2 - 4 mV. **b**, Small-bias dI/dV spectra (open dots) showing peaks inside the superconducting gap due to YSR states. Dashed lines over each spectrum are fits obtained using a numerical deconvolution method (see Methods). Spectra are shifted vertically for clarity. Stabilization parameters for STS: $V_{stab}$ = 2.5 mV, $I_{stab}$ = 100 pA, $V_{mod}$ = 20 μV. The corresponding deconvoluted LDOS is plotted in **c**. The green curve in **c** corresponds to the reference substrate density of states.

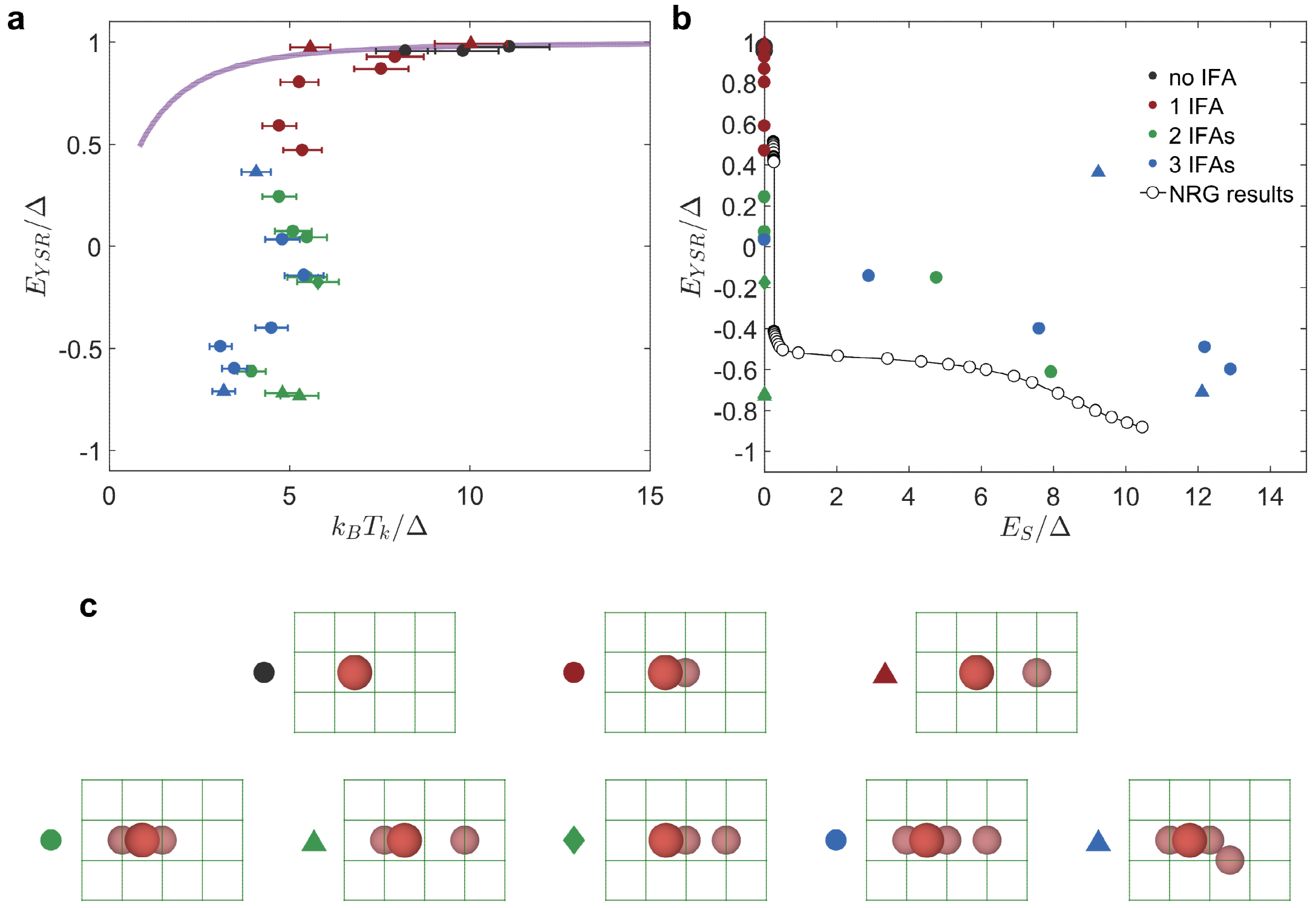

**Figure 3 | Correlation of Kondo temperature, exchange splitting, and YSR state energy.** **a**, Scatter plot of the energy of the larger YSR peak versus the Kondo temperature, both extracted from the experimentally investigated Fe adatom with different IFA assemblies. Various symbols in the plot represent various types of IFA assemblies as indicated in **c**. Horizontal error bars indicate the estimated errors in the fitting of the Kondo-resonance width. The violet curve in **a** represents the slave-boson mean-field theory calculation. **b**, Scatter plot of the energy of the larger YSR peak versus the exchange splitting of the Kondo resonance, both extracted from the experimentally investigated Fe adatom with different IFA assemblies (assignment of the symbols to the different assemblies, see **c**). Results of the NRG calculations, which are plotted as open black dots, showing qualitative agreement with the experiments. All energy scales are normalized to the energy gap (Δ) of the superconducting substrate.

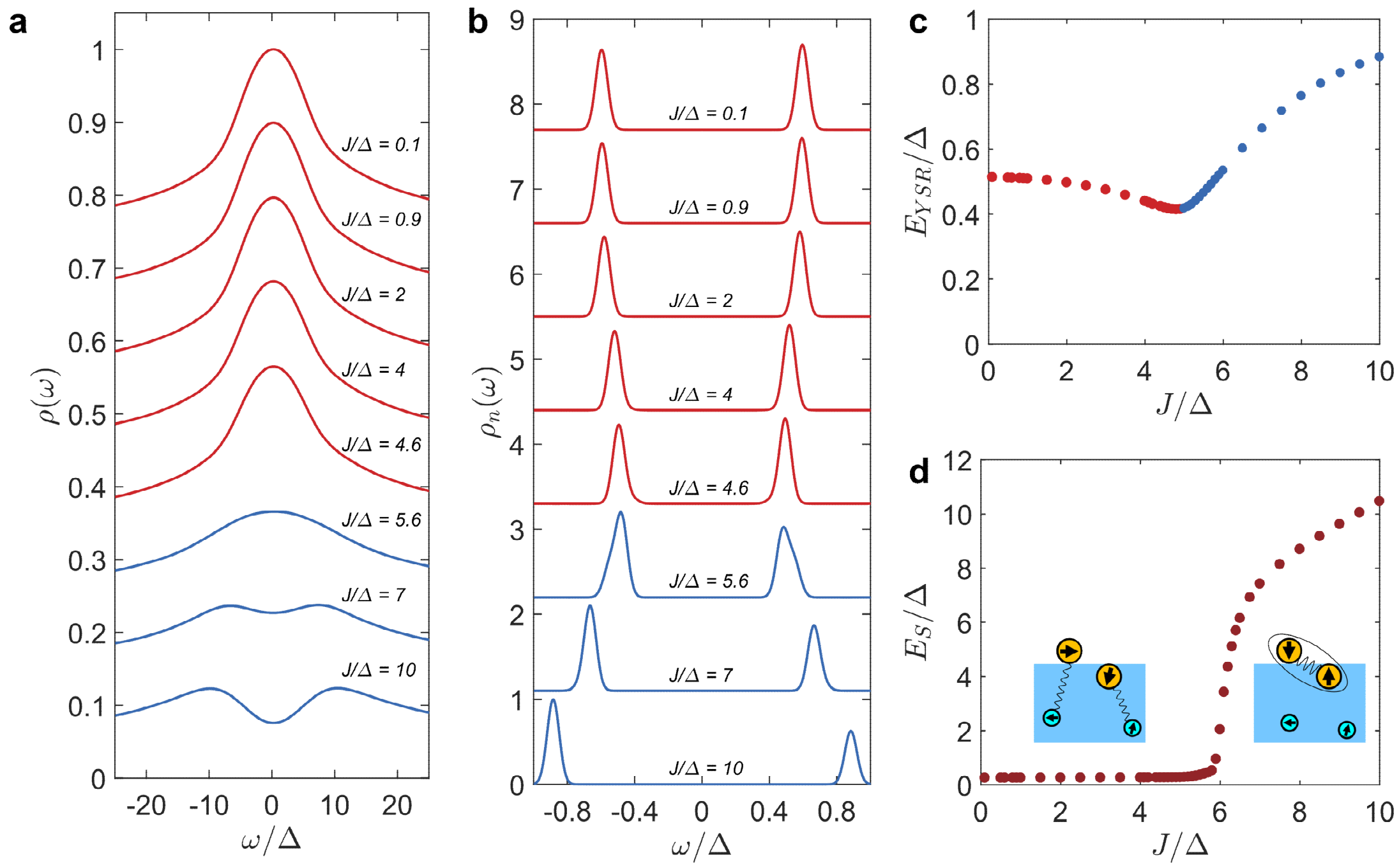

**Figure 4 | Numerical renormalization group calculations. a**, Spectral function of the impurity representing the Fe adatom, which is magnetically coupled to a second impurity representing the IFA assembly, on the energy scale of the Kondo peak for various magnetic interaction strengths *J*. The superconducting gap is washed out here using a convolution which mimics the experimental lock-in-modulation broadening. Red and blue colors indicate the nature of the ground state of the system (Kondo vs. AFM) controlled by the strength of *J*. **b**, Discrete (sub-gap) part of the spectral function; a narrow Gaussian kernel is used to broaden the delta peaks. **c**, Energy of the YSR state as a function of *J*. **d**, Width of the exchange-splitting gap as a function of *J*. The insets illustrate the two regimes of decoupled Kondo singlets (small *J*) and AFM dimer (large *J*).

## Supplementary Figures

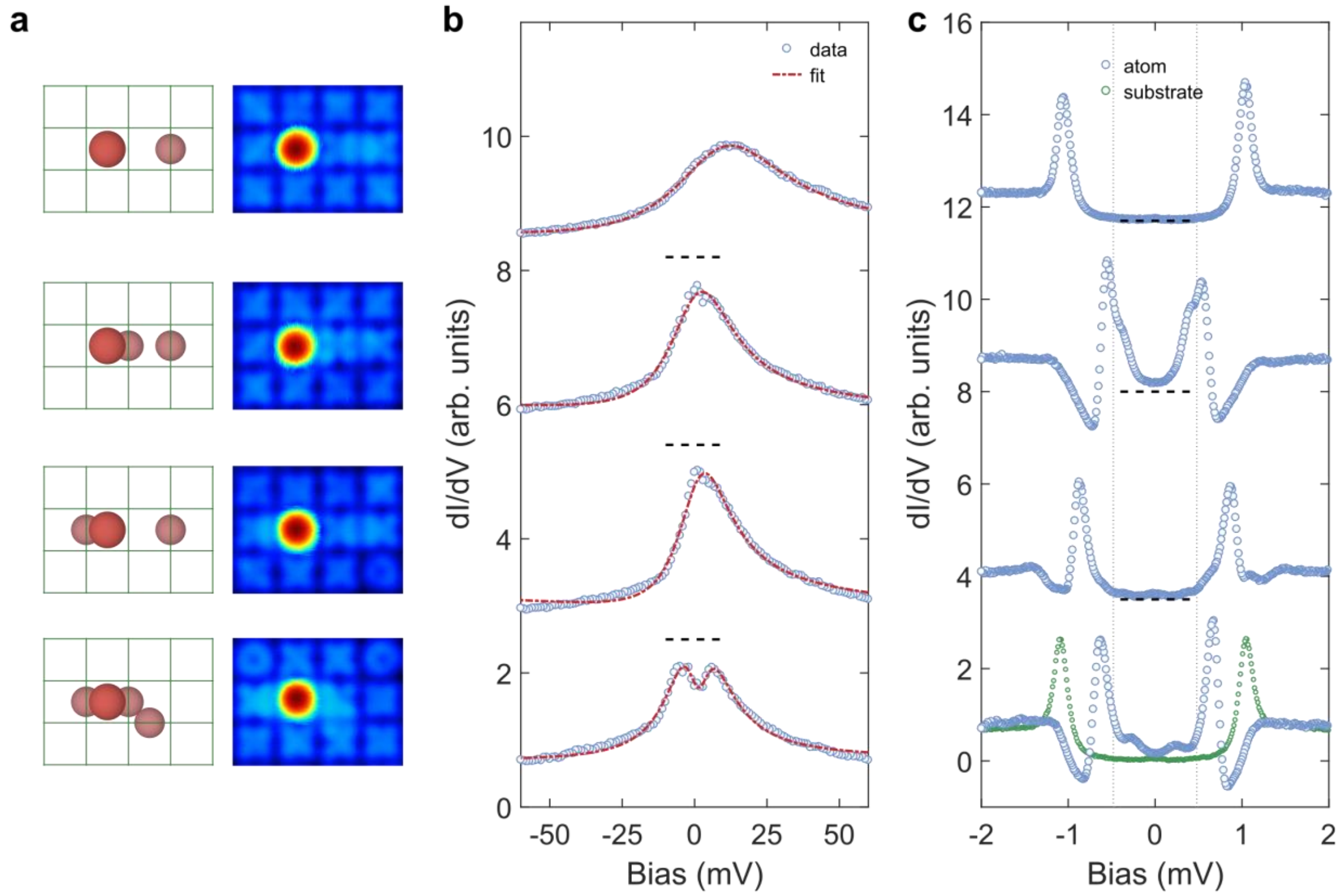

**Supplementary Figure 1 | dI/dV spectroscopy on Fe adatoms of additional Fe clusters.** **a,** Schematic top view diagram and corresponding STM images of different clusters of Fe atoms. **b**, Large bias range dI/dV spectra (open dots) showing the Kondo resonance. The dashed lines are fits to a Fano function, which is multiplied by a Lorentzian dip (see Methods). Stabilization parameters for STS: $V_{stab}$ = 100 mV, $I_{stab}$ = 100 pA, $V_{mod}$ = 2 - 4 mV. **c**, Small bias range dI/dV spectra measured with a superconducting tip (open dots), showing the peaks due to the YSR states. Stabilization parameters for STS: $V_{stab}$ = 2.5 mV, $I_{stab}$ = 100 pA, $V_{mod}$ = 20 µV.

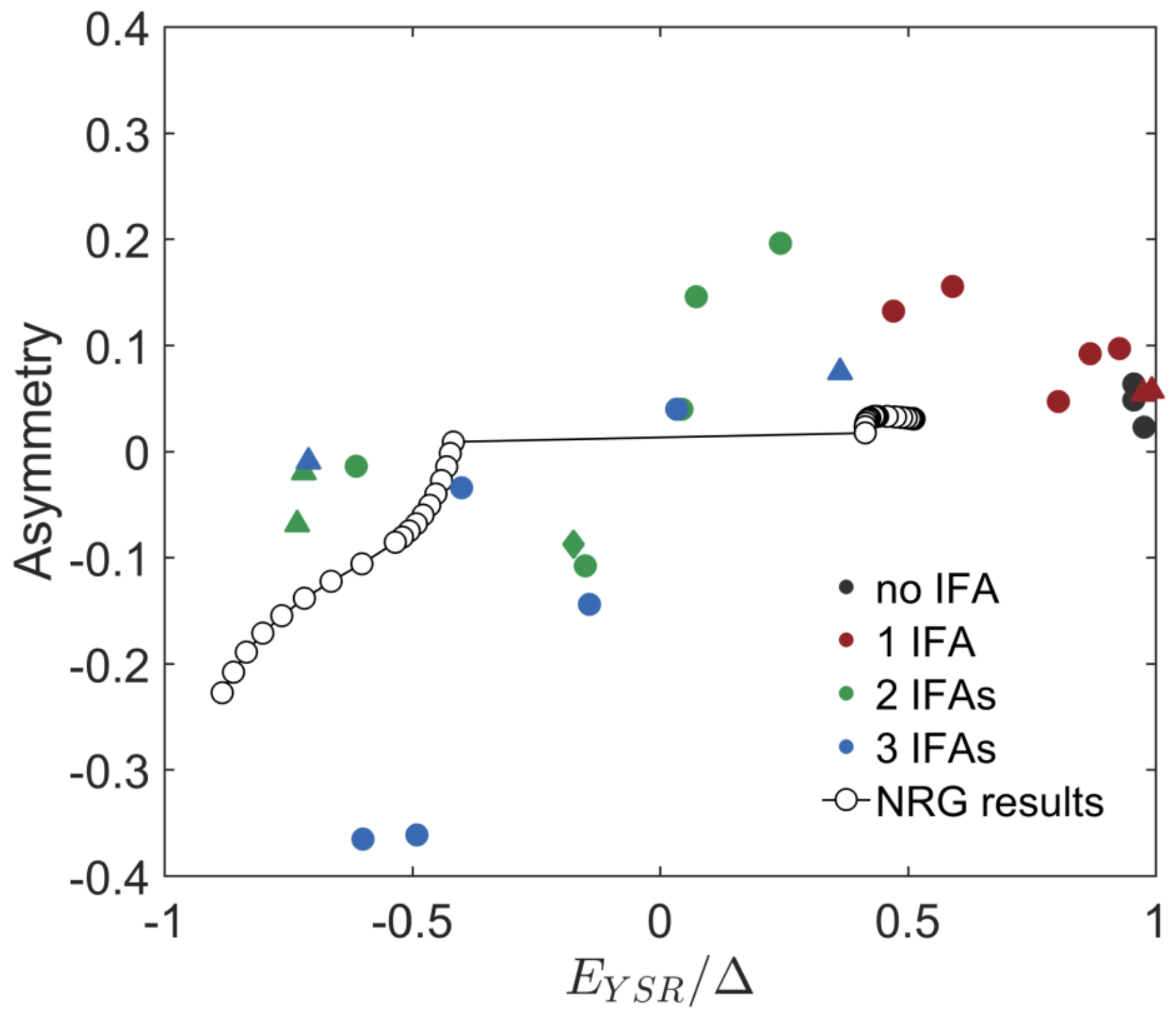

**Supplementary Figure 2 | Asymmetry of the YSR peaks as a function of bound state energy.** We observe that the intensities of the YSR peaks are asymmetric and as a general trend reverse sign with increasing number of IFAs in the assembly. To characterize this effect, we define the asymmetry as:

$$Asymmetry = \frac{h_R - h_L}{h_R + h_L},$$

where $h_R(h_L)$ is the amplitude of the YSR peak on the positive (negative) bias side. The corresponding experimental values are shown as filled symbols, where different symbols represent the various Fe clusters defined in Figure 3c of the main manuscript. The plot also shows the results of the NRG calculations as open dots for comparison, showing qualitative agreement with the experimental trend.

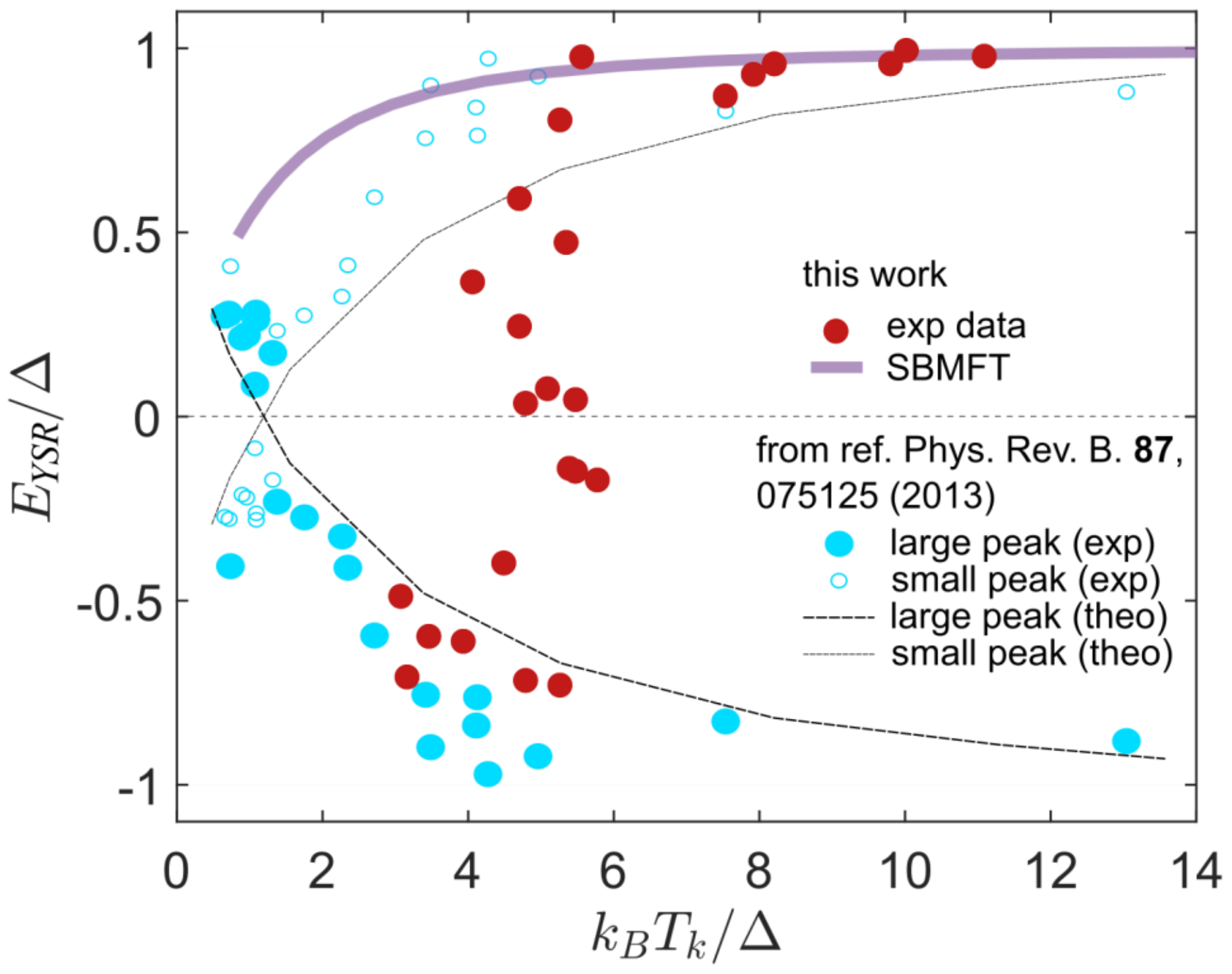

**Supplementary Figure 3 | Comparison of YSR energy versus Kondo temperature with literature.** Blue data points and black lines are adapted from the experimental data and numerical renormalization group (NRG) calculations, respectively, for the system of MnPc molecules on Pb(111)[1]. Brown filled circles and violet line are our experimental data points and Slave boson mean-field theory (SBMFT) calculations, respectively, for the system of Fe clusters on TaO.

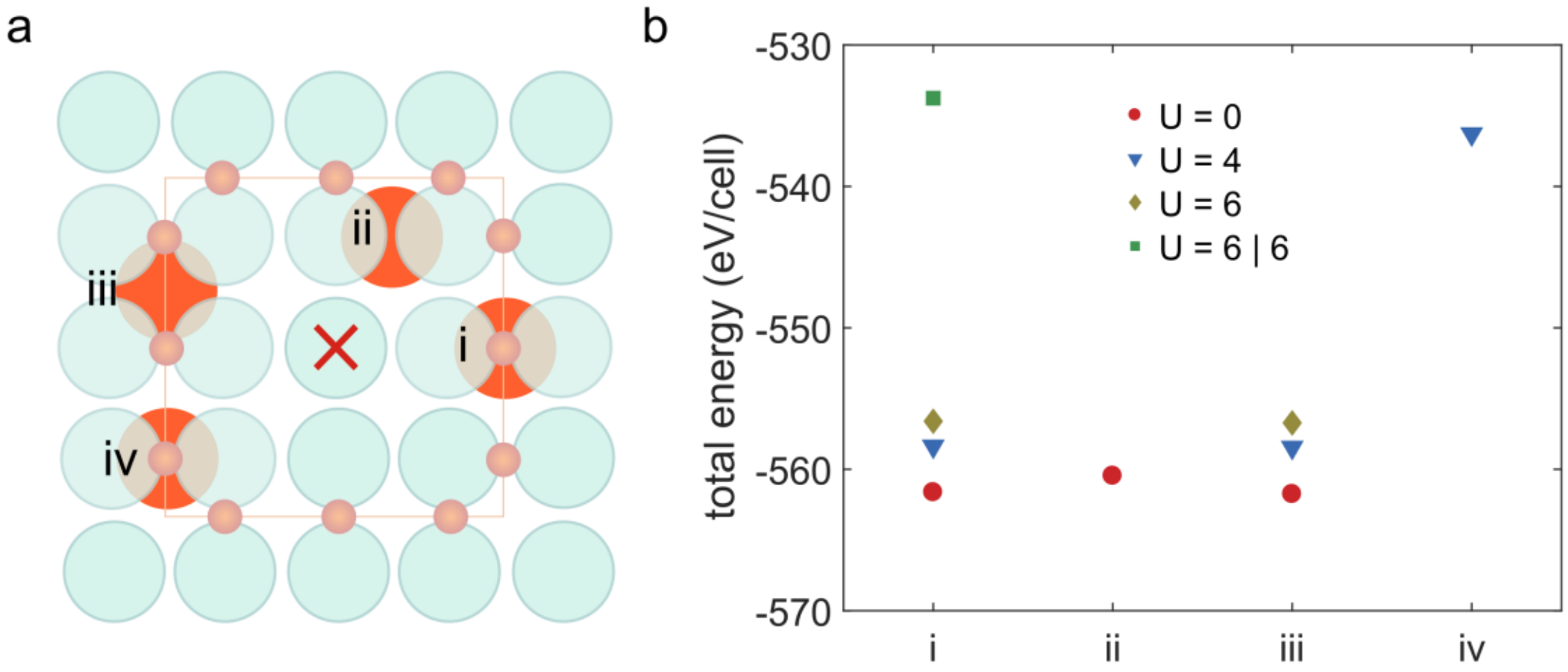

**Supplementary Figure 4 | Interstitial Fe atom (IFA) within DFT approach. a,** Suggested positions with one IFA per (3 x 3) plaquette (initial sites) that are marked by letters (i-iv). **b,** Total energy comparison per unit cell as obtained within GGA+U ($U_{Fe\text{-}d}$ | $U_{O\text{-}p}$) simulations and compared between different compositions (i-iv). We have also considered an additional position of an IFA in a hollow cite of the Ta island. The latter configuration leads to significant reconstruction of the surface and a loss of the (3 x 3) plaquette structure, so we dismissed it. Among the other positions, one can see that, after relaxation, the IFA that was initially located on (iii) converged to location (i). In this setup, the IFA occupies approximately the center of the Ta octahedra with irregular base and two apexes, connected by the bridge oxygen atom. Among all considered configurations, only case (i) satisfies the criteria of lowest energy and smallest distortion of the surface (3 x 3) plaquette structure. To control the charge localization in the system, we used an on-site Coulomb parameter for Fe and O states, which was found to improve the convergence of the electronic and ionic minimizations significantly but did not change the structure qualitatively. The main surface change, induced by the presence of such an IFA in the first interlayer, is an elevated oxygen atom in the bridge position together with a slight upwards shift of the nearest Ta atoms. This change would be reflected in an STM image as an enhancement of the cross-like shape of the (3 x 3) plaquette near the IFA[2], which is indeed seen in the experiment (see Fig. 1a of the main manuscript).

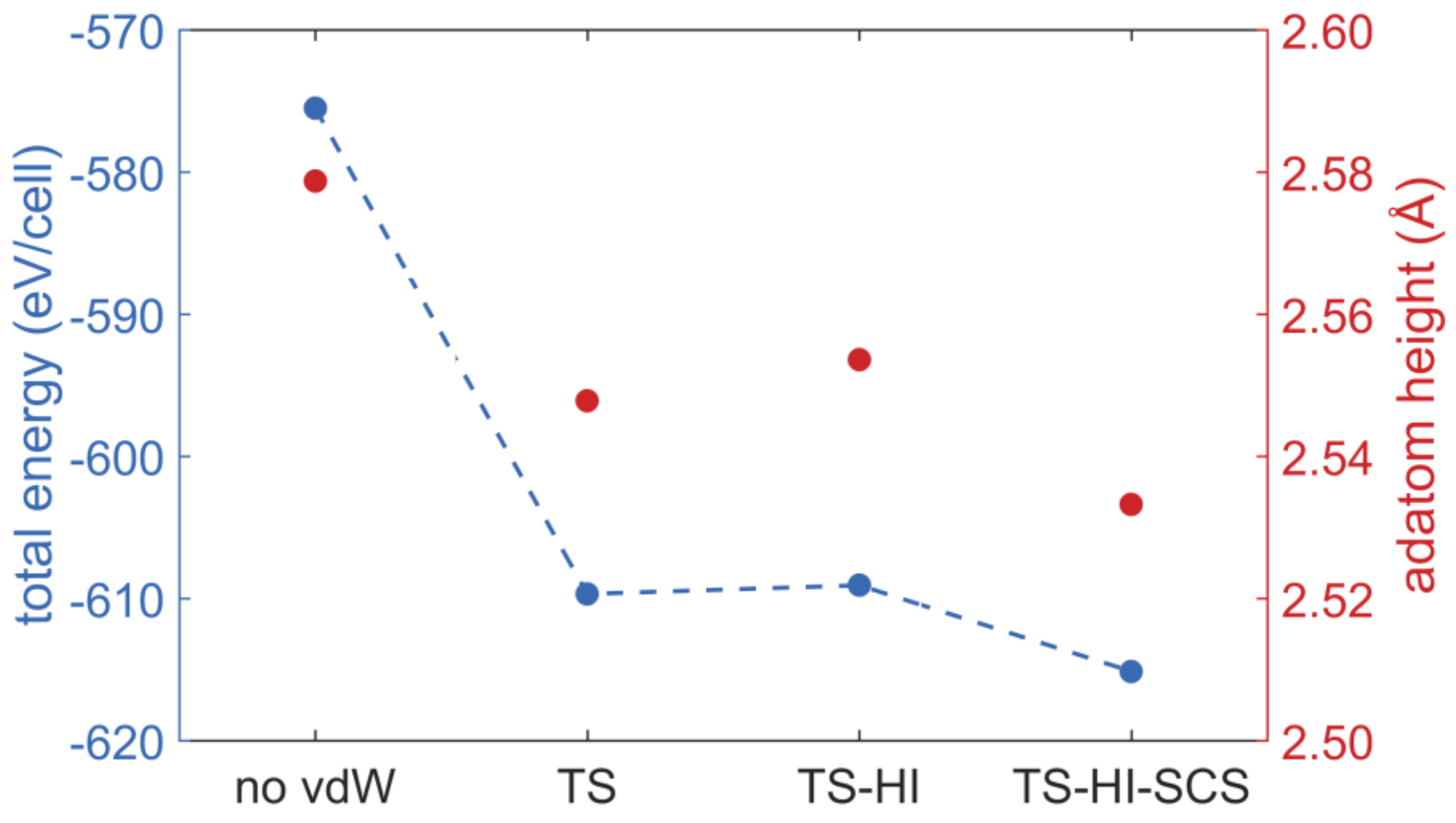

**Supplementary Figure 5 | VdW approximations for the Fe adatom at the center of the (3 x 3) plaquette.** Comparisons of the heights and total energies (in the slab with two surfaces) for the system of one Fe adatom at the center of the (3 x 3) plaquette with respect to different vdW frameworks, implemented in the VASP package: no vdW (no van-der-Waals interaction), TS (Tkatchenko-Scheffler method), TS/HI (TS scheme with iterative Hirschfield partitioning) and TS/HI-SCS (self-consistent scheme of TS/HI). The Fe adatom was allowed to move only along the *z*-direction, while the two in-plane adatom coordinates were fixed to the center position resulting from the full free relaxation of the surface within the TS/HI scheme. The reduction of the repulsive interaction between the adatom due to the dispersion forces, acting between the adatom and the polarizable substrate, is clearly visible here. This effect indicates the presence of a strong dynamical dipole-dipole interaction, that should be accounted for in the adsorption process on such surfaces, as it was predicted earlier[2]. Within a full surface relaxation, the screening TS/HI-SCS method or switching the vdW part off (no vdW) lead to the relaxation of the adatom to the nearest hollow position, which is off-center. We attribute this error to a rather effective way of the inclusion of dynamics of the interaction between fluctuating charges as part of a pseudopotential and its possible reduction in the DFT-vdW-SCS method. The obtained height of the Fe adatom over the surface is approx. 2.5 Å (TS/HI), that fits to the experimental results[3,4].

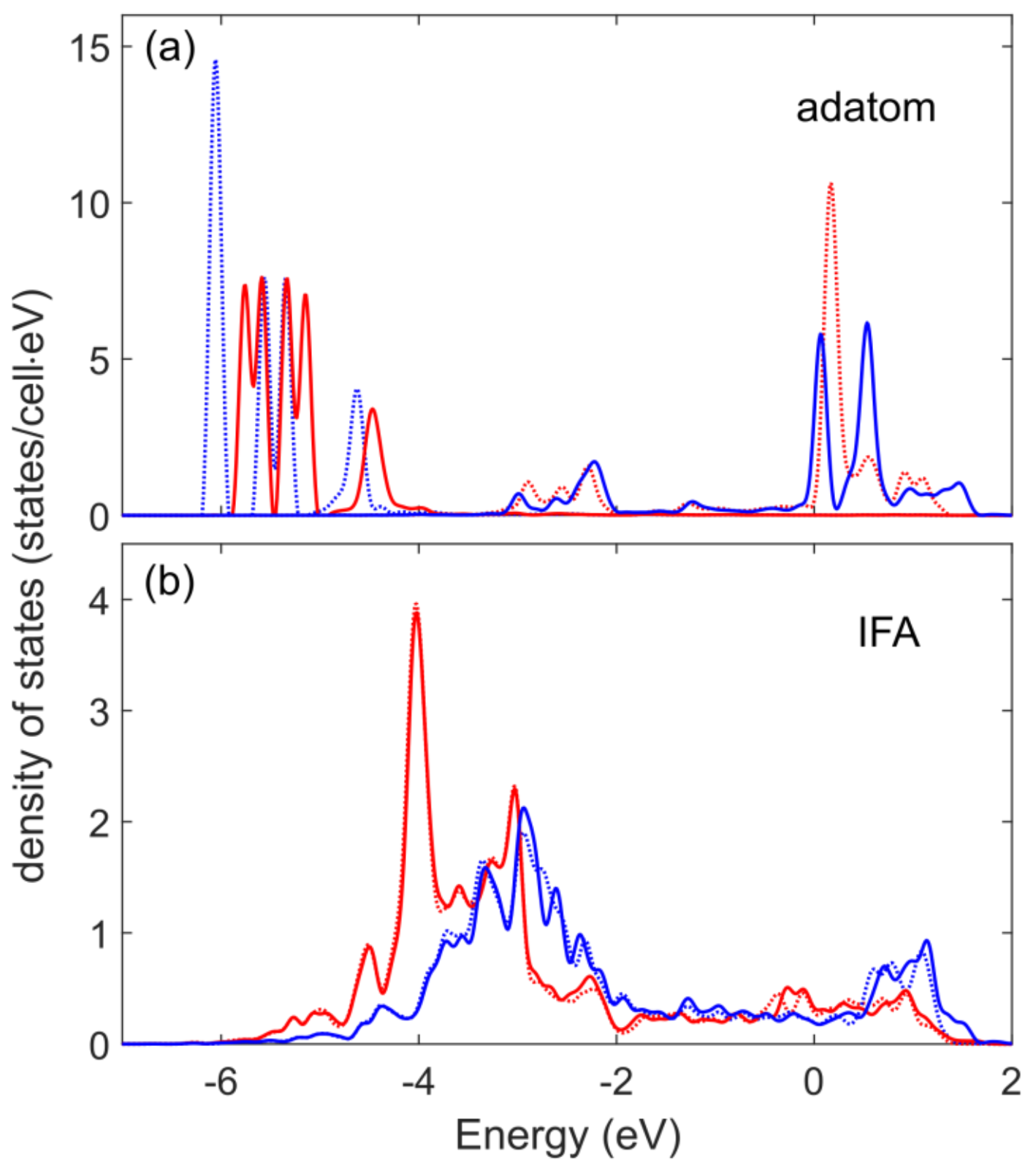

**Supplementary Figure 6 | Magnetization of the Fe adatom in the presence of an IFA.** Density of states for the *d*-states of the Fe adatom (**a**) and the IFA (**b**) in the slab with both Fe atoms. Here, continuous lines refer to the slab with 2 atoms (Fe adatom and IFA) and dashed lines refer to the slab with one Fe atom in absence of the other (either adatom or IFA). Red and blue colors denote spin-up and spin-down densities, respectively. The position of the adatom was extracted from the vdW calculations**,** while the IFA occupies the space beneath the bridge oxygen atom (case (i) in Supplementary Figure 4). The system is relaxed within the spin-polarized GGA+U scheme with the same parameters used for the simulations with the IFA. The distance between the IFA and the adatom is 6.5 Å after relaxation. One can see that the IFA has a small but non-zero magnetic moment (0.9 $\mu_B$) which does not change upon removal of the Fe adatom. In contrast, the Fe adatom has a large magnetic moment of 3.1 $\mu_B$, and a splitting of the peaks near the Fermi energy appears due to the magnetic interaction with the IFA.

## Supplementary references